\documentclass[pdftex,twocolumn,epjc3]{svjour3}          
\usepackage{mathtools}
\usepackage{lineno}
\RequirePackage[T1]{fontenc}
\usepackage{array}
\usepackage[normalem]{ulem} 
\smartqed  

\RequirePackage{graphicx}
\usepackage{amssymb}
\RequirePackage{mathptmx}      
\RequirePackage{flushend}
\RequirePackage[numbers,sort&compress]{natbib}
\RequirePackage[colorlinks,citecolor=blue,urlcolor=blue,linkcolor=blue]{hyperref}

\journalname{Eur. Phys. J. C}

\begin{document}
\title{End-to-end reconstruction of ultra-high energy particle observables from radio detection of extensive air showers}

\author{Kewen Zhang\thanksref{e1,addr1,addr2} \and
        Kaikai Duan\thanksref{addr1,addr2} \and
        Ramesh Koirala \thanksref{addr3,addr4,addr5} \and
        Matías Tueros \thanksref{addr6,addr7} \and
        Chao Zhang\thanksref{e2,addr3,addr4} \and
        Yi Zhang\thanksref{addr1,addr2} 
}

\thankstext{e1}{e-mail: kwzhang@pmo.ac.cn}
\thankstext{e2}{e-mail: chao.zhang@nju.edu.cn}

\institute{Key Laboratory of Dark Matter and Space Astronomy, Purple Mountain Observatory, Chinese Academy of Sciences, No. 10 Yuanhua Road, Nanjing, China\label{addr1}   \and
          School of Astronomy and Space Science, University of Science and Technology of China, Hefei 230026, China\label{addr2}   \and
          School of Astronomy and Space Science, Nanjing University, Nanjing 210023, China\label{addr3}    \and
          Key laboratory of Modern Astronomy and Astrophysics, Nanjing University, Ministry of Education, Nanjing 210023, China\label{addr4}    \and
          Space Research Centre, Faculty of Technology, Nepal Academy of Science and Technology, Khumaltar, Lalitpur, Nepal \label{addr5} \and
          Instituto de Física La Plata - CCT La Plata - CONICET, Diagonal 113 y 63, La Plata, Argentina\label{addr6} \and
          Depto. de Física, Fac. de Cs. Ex., Universidad Nacional de La Plata, Calle 48 y 115, La Plata, Argentina \label{addr7}
}


\maketitle

\begin{abstract}
The radio detection of very inclined air showers offers a promising avenue for studying ultra-high-energy cosmic rays (UHECRs) and neutrinos. Accurate reconstruction methods are essential for investigating the properties of primary particles. Recently, we developed an analytical least-squares method to reconstruct the electric field using three polarization components. The reconstruction yields no bias, with a 68\% confidence interval of [-0.02, 0.02], and a standard deviation of 0.04. Using this reconstructed electric field, we perform a realistic reconstruction of the the properties of primary particles. We employ a spherical wave model combined with an angular distribution function (ADF) for arrival direction reconstruction, achieving an angular resolution of 0.04$^\circ$. This paper also presents an energy reconstruction in which we account for the effects of geosynchrotron radiation in inclined air showers, we implement an air density correction in the energy reconstruction, resulting in a 10\% resolution in energy estimation. These findings demonstrate the reliability and effectiveness of our reconstruction methodology, paving the way for future detection experiments using sparse antenna arrays.

\end{abstract}

\tableofcontents

\section{Introduction}
\label{sec:intro}
Extensive air showers (EAS) are generated by the bombardment of cosmic rays on the nuclei of atoms in the Earth atmosphere. EAS have been successfully detected using several types of detectors in the last few decades, which offered valuable information to investigate the origin, the propagation mechanisms and properties of high energy particles \cite{Gaisser:2016uoy}. The cosmic ray flux decreases with energy following a steep power-law $\varphi(E) \propto E^{-\gamma} $ with the spectral index $\gamma$ ranging between 2.7 and 3.3 in the ultra-high energy domain. \cite{TelescopeArray:2015dcv,PierreAuger:2018pmw}. Consequently, detecting ultra-high-energy particles above 1 EeV requires large detector arrays capable of observing inclined air showers with zenith angles  $\theta > 60^\circ $, such as the Pierre Auger Observatory \cite{PierreAuger:2018pmw, PierreAuger:2014jss}. Inclined air showers are particularly important due to their potential to probe Earth-skimming ultra-high-energy neutrinos, which can produce extensive air showers upon interacting near the Earth’s surface. Due to their large signal footprint on the ground, this type of showers are an ideal target for experiments utilizing radio detection techniques \cite{Huege:2023pfb, GRAND:2018iaj, Zeolla:2023phg}.

Radio signals are produced when the charged particles of the EAS propagate through the atmosphere. the dominant contribution to the radio signal comes from the geomagnetic effect \cite{Kahn:1966}, where charges are deflected by the Earth’s magnetic field, generating a time-varying transverse current. A subdominant contribution arises from the Askaryan effect, which results from the charge excess that develops at the shower front as the particle cascade progresses\cite{Askaryan:1962, Aab:2014}. In EAS, the relative strength of the Askaryan effect is typically below 20\%, as summarized in \cite{frank:2016, Horandel:2019qwu}. For instance, measurements from AERA indicate that the Askaryan contribution accounts for (14 $\pm$ 2)\% of the total field strength \cite{Aab:2014}. 
The polarization of the electric field from the geomagnetic effect follows the direction of the Lorentz force, resulting in linear polarization in the shower plane (the plane perpendicular to the shower direction of motion). In contrast, the Askaryan effect produces radially polarized fields directed toward the shower axis. At any given observation point, the total polarization is the vector sum of these two contributions. The signal strength and direction depend on the observer’s position relative to the shower axis.
Both emission mechanisms exhibit a characteristic ring-like structure in the shower plane due to temporal compression of the radio signal, known as the Cherenkov ring. The interplay between the geomagnetic and Askaryan components can either amplify or suppress this pattern, influencing the observed signal morphology \cite{Schluter:2020tdz}.

An often unnoticed effect originating from geosynchrotron radiation has recently gained interest. Inclined air showers ($\theta > 60^\circ$) differ notably from that of vertical air showers, as their shower maximum $X_\mathrm{max}$ occurs higher in the atmosphere. Due to longer propagation in lower-density air, electrons and positrons can gyrate more in the geomagnetic field before attenuation, causing the radio emission to enter the geosynchrotron regime \cite{James:2022, geosynchrotron2024}, where both geomagnetic and Askaryan contributions are modified and geosynchrotron radiation becomes significant.
An evident decrease in geomagnetic radiation in inclined air showers developed in low air density is shown in \cite{geosynchrotron2024}. Earlier indications of zenith dependence were observed by LOFAR, where the relative strength of the Askaryan effect was found to be (20.3 ± 1.3) \% for near-vertical air showers at 225 m, and (3.3 ± 1.0) \% for inclined air showers at 25 m \cite{Schellart:2014oaa} from the shower core. This value decreases to around 2\% for inclined air showers with zenith angle $\theta = 80^\circ$ for the magnetic field at the Dunhuang, China (site of a prototype with 300 antennas of the Giant Radio Array for Neutrino Detector (GRAND), called GRANDproto300 ) \cite{Chiche:2022ppi}.  Additionally, for the early and late observers, the different paths to the shower maximum X$_\mathrm{max}$ lead to the early-late effect \cite{Schluter:2020tdz}, which makes another asymmetry of radio signal from inclined air showers. To investigate the general structure of the radio footprint, the antennas to detect vertical air showers are primarily concentrated within a compact region forming a dense antenna array, which presents limitations to detect inclined air showers that produce signal footprints covering a much larger area. To study this type of events, sparse antenna arrays are needed. Due to all these differences between the signal characteristics of vertical and inclined showers, the standard reconstruction methods developed for vertical air showers are not well suited for inclined air showers, and the development of new methods tailored for inclined showers becomes indispensable. 

To address this need, we implement a dedicated reconstruction pipeline optimized for inclined geometries. We apply a novel analytical chi-square solution for the reconstruction of the electric field from the measured voltages at the antenna. This reconstructed electric fields are then used with the ADF method \cite{Decoenethesis:2020,decoene2021reconstruction} for determining the arrival direction of the primary particle and the radio emission point, related to the shower maximum \cite{decoene2023radio}. Combining this with the air density correction, a determination of the energy of the primary particle initiating the cascade is performed. 


%


In the following sections, we begin by presenting a library of cosmic rays events with a simulated voltage signal including galactic noise, which serves as the starting point for this investigation.  Then we  scrutinize the electric field reconstruction process, as it forms the foundation for reconstructing other physical properties.
Subsequently, we present the methods and results of the reconstructions for direction and energy, and the performance of the methods. Finally, we summarize our results and discuss the uncertainties.

\section{Air shower library}
\label{sim_set}

\begin{table}
    \centering
    \begin{tabular}{l|l} 
        \hline
        \textbf{Number of showers}  & 4160 \\ \hline
        \textbf{Primaries} & p [50\%], Fe [50\%] \\ \hline
        \textbf{Energies E / $\mathrm{eV}$}  & 10$^{17.1}$ , 10$^{17.2}$ , ... , 10$^{18.6}$ [24 steps] \\ 
        \hline
        \textbf{Zenith angles}  & 63.0$^\circ$ , 67.8$^\circ$ , ... , 87.1$^\circ$  \\ 
        \hline
        \textbf{Azimuth angles}  & 0$^\circ$ , 45$^\circ$ , ... , 180$^\circ$ \\ 
        \hline
        \textbf{Hadronic model}  &  SIBYLL23d \\ 
        \hline
        \textbf{Thinning $\epsilon_{\mathrm{thin}}$} & 1 $\times 10^{-5}$ \\ 
        \hline
        \textbf{Atmospheric model} & Extended Linsley's std. atmosphere \\ \hline
        \textbf{Geomagnetic field} & \begin{tabular}[c]{@{}l@{}}Intensity: 55.997 $\mu$T \\ Inc: 60.79$^\circ$  Dec: 0.36$^\circ$\end{tabular} \\ \hline
        \textbf{time bin} & 0.5ns \\ \hline
        \textbf{time window} & 1000ns \\ \hline
    \end{tabular}
    \caption{Parameters of the air shower library generated by the ZHAireS simulations.}
    \label{tab:my_label}
\end{table}

To perform this study we used the ZHAireS package \cite{ZHAireS2012} to make a library of EAS and their induced radio signals. The zenith angle was defined in $\mathrm{log10}(1/\mathrm{cos}(\theta)) = 0.08 $ steps. The range of this and other relevant shower parameters is presented in Table \ref{tab:my_label}. The atmospheric index of refraction was modeled using an exponential function, with scale height of 8.2km and a refraction index of 1.000325 at sea level.

The magnetic field is set to be the local magnetic field strength of the Dunhuang site of GRAND, which is 1100 meters above the sea level. In total, this simulation set contains 4160 showers split equally between protons and Iron as primaries. 
On each event the signal is simulated on 160 antenna positions organized into a star-shape pattern with 8 arms, 20 antennas per arm, with a radius adjusted to properly sample the Cherenkov ring. 16 additional antennas were randomly distributed at ground level. The simulated electric field is then used to compute the antenna response, and get an open circuit voltage at the antenna connector. We test the three-polarization HORIZON antenna, identical to that in \cite{Zhang:2025qef} but placed 0.2 m lower than in the GP300 design \cite{GRANDlib}. Positioned 3 m above ground, it covers the 50–200 MHz band relevant for EAS radio emission. 

The  only noise source considered in the simulation library is galactic noise, that is the primary noise source, especially at the lower part of the frequency band. This noise is generated using the LFmap model \cite{Lfmap:2007, Zhang:2025qef} for 18hs local sidereal time (LST), that is considered to be the worst case scenario. Since the electronic and other environmental noise sources are strongly dependent on the experimental setup and the experiment location, they are not included in this work. The simulated electric field is decomposed into three orthogonal polarization components: Ex (South-North), Ey (East-West), and Ez (Down-Up).

\begin{figure*}[ht!]
    \centering
    \begin{minipage}{0.45\textwidth}
        \centering
        \includegraphics[width=\textwidth]{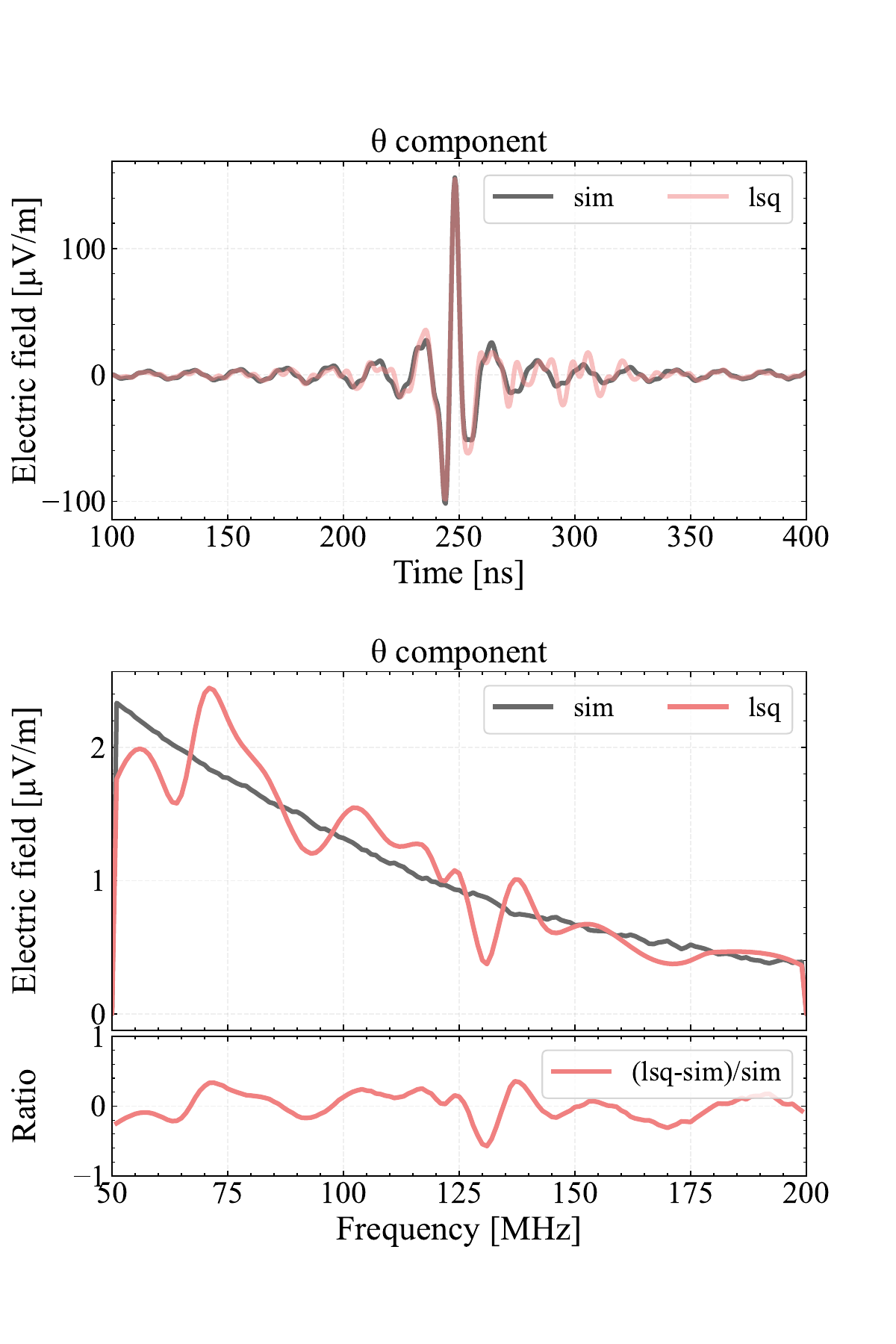}
    \end{minipage}%
    \hspace{0.05\textwidth}
    \begin{minipage}{0.45\textwidth}
        \centering
        \includegraphics[width=\linewidth]{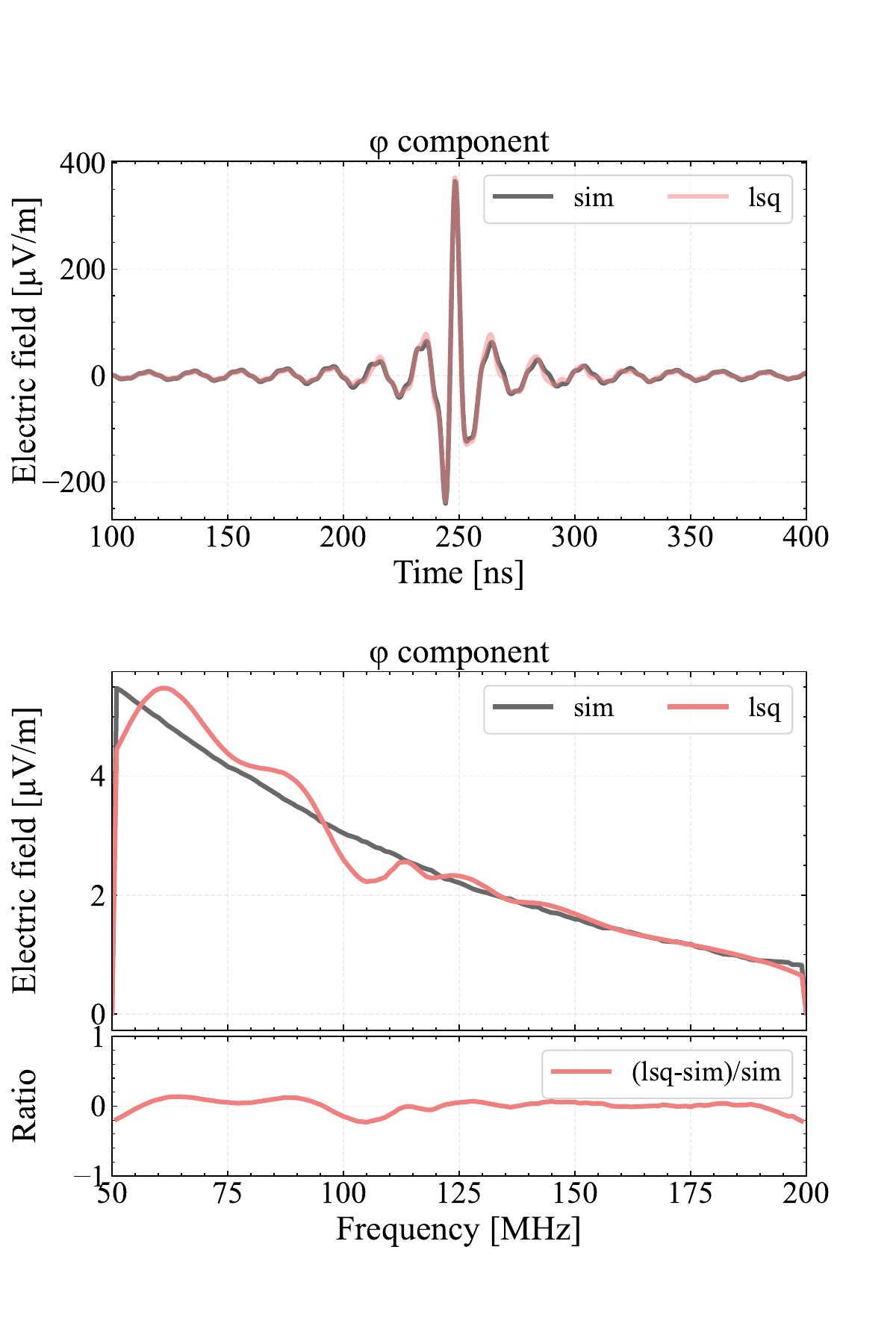}
    \end{minipage}
    \caption{Reconstructed electric field using the least squares (lsq) method, displayed in both the time domain (top) and frequency domain (bottom). The left panel shows the $\theta$ component, while the right panel shows the $\varphi$-component of the electric field. This trace is from a proton-induced air shower with a primary energy of 0.631EeV, zenith angle of $67.8^\circ$, and azimuth angle of $45^\circ$.}
    \label{fig:e_field_rec}
\end{figure*}
\section{Electric field reconstruction}

The detected radio signal contains the amplitude, polarization and frequency spectrum of the incomming electric field convoluted with the response of the antenna.  The reconstruction of the electric field from the measured voltage signals is a critical step, as it is from the electric field that other properties of the cascade can be calculated. One example of this is the energy fluence, that allows the determination of the primary energy. 


Adding a vertical arm improves the detection of inclined air showers for some incoming directions while maintaining the sensitivity to vertical ones, but more importantly, it allows for an unambiguous determination of the electric field polarization direction. This helps to reject background noise containing unexpected polarization patterns, which is usually the case for anthropogenic sources. On the down side, adding the vertical polarization introduces complexity to the reconstruction of the electric field. 

Most of the existing methods for the reconstruction of the electric field are designed to handle the signals received by antennas with two polarizations.  The matrix inversion method, for example, works well with a 2×2 response matrix if the noise is ignored, but it struggles when three polarizations are involved \cite{Zhang:2025qef}. For in-ice experiments with compact sub-stations each containing several antennas, a forward folding technique has been developed using frequency slopes in the least-squares ($\chi^2$) method,  which  yields better results than the matrix inversion method when SNR < 10 \cite{Glaser:2019}. Inspired by this method, we have developed an analytical least-squares ($\chi^2$) method \cite{Zhang:2025qef} which does not require any prior assumptions about the signal characteristics (amplitude, frequency slope, phase) and provides a precise and straightforward reconstruction. This work is the first application of this new method to the reconstruction of shower observables.

\subsection{Analytical least-squares method}

Reconstructing the electric field is simpler in the frequency domain by using the Fourier transform of the measured antenna voltage $\mathcal{V}(f)$. When noise $\mathcal{N}(f)$ is added to the open-circuit voltage, the relationship between the measured antenna voltage and electric field $\mathcal{E}(f)$ can be written as

\begin{equation}
        \boldsymbol{\mathcal{V}}(f) = 
    \begin{pmatrix}
        \mathcal{H}_{1, \theta} & \mathcal{H}_{1, \varphi}\\
        \mathcal{H}_{2, \theta} & \mathcal{H}_{2, \varphi} \\
        \mathcal{H}_{3, \theta} & \mathcal{H}_{3, \varphi}
    \end{pmatrix}
    \begin{pmatrix}
        \mathcal{E_\theta}\\
        \mathcal{E_\varphi}
    \end{pmatrix}+\mathcal{N}(f),
    \label{eq:v_matrix_simp}
\end{equation}

where $\mathcal{H}(f)$ is the antenna response matrix with subscripts 1, 2 and 3 representing the three antenna arms, while $\theta$ and $\varphi$ represent the electric field components in spherical coordinates.  Since the far-field approximation is employed, the small $r$-component along the direction of signal propagation is neglected. The electric field is expressed as $\boldsymbol{\mathcal{E}}=\left(\begin{array}{l}\mathcal{E}_{\theta} \\ \mathcal{E}{_\varphi}\end{array}\right)$.

Given a certain signal arrival direction, the known response matrix and noise distribution, the $\chi^2$ can be defined as:


\begin{equation}
\begin{multlined}
\chi^2  =  \sum_{i=1}^3 \left( \frac{\mathbf{\mathcal{V}}_i - \mathcal{H}_i \left( \begin{array}{l} \mathcal{E}_\theta \\ \mathcal{E}_\varphi  \end{array} \right) } {\sigma_{\mathcal{V}_i}}\right)^2 \\
 = (\boldsymbol{\mathcal{V}} - \boldsymbol{\mathcal{H}} \boldsymbol{\mathcal{E}})^T \sigma_{\mathcal{V}}^{-1} (\boldsymbol{\mathcal{V}} - \boldsymbol{\mathcal{H}} \boldsymbol{\mathcal{E}})
\end{multlined},
    \label{eq:v_matrix}
\end{equation}

where $i$ represents the i-th polarization, and the covariance matrix  $\sigma_\mathcal{V}=\operatorname{diag}\left(\sigma_{\mathcal{V} 1}, \sigma_{\mathcal{V} 2}, \sigma_{\mathcal{V} 3}\right)$ is given by the spectrum of the noise. 

An analytical solution of the electric field is obtained by minimizing $\chi^2$ \cite{Behnke:2013}, and is given by
\begin{equation}
    \boldsymbol{\mathcal{E}} = (\boldsymbol{\mathcal{H}}^T \sigma_{\mathcal{V}}^{-1} \boldsymbol{\mathcal{H}})^{-1} \boldsymbol{\mathcal{H}}^T \sigma_{\mathcal{V}}^{-1} \boldsymbol{\mathcal{V}}.
    \label{eq:e_inverse}
\end{equation}

This equation indicates that with accurate knowledge of the background noise and the antenna response, the electric field can be reconstructed from the voltages measured at each antenna.

\subsection{ Electric field resolution}  


\begin{figure}[htbp!]
\centering
\includegraphics[width=0.45\textwidth]{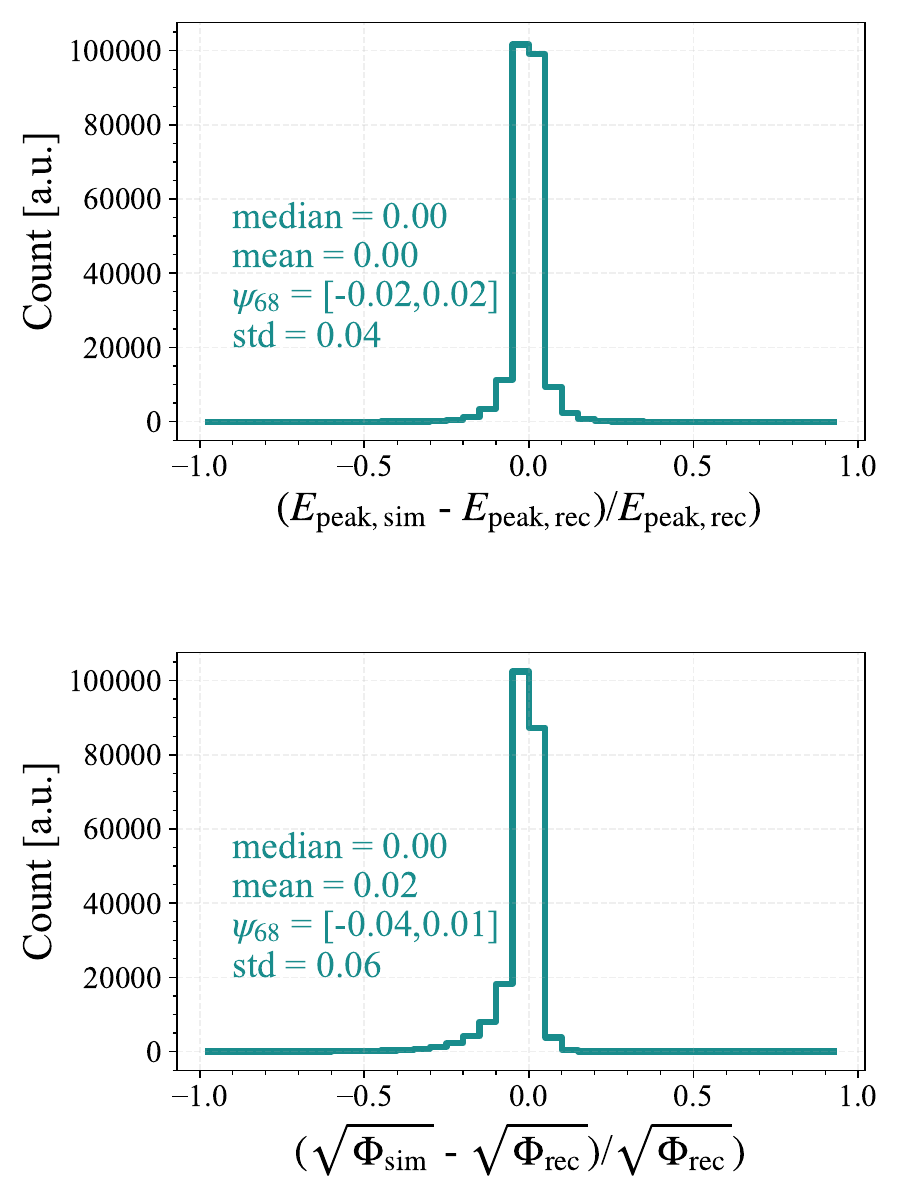}
    \caption{Distribution of the relative error in the reconstructed electric field compared with Monte Carlo simulations. Top: the  maximum (peak) of the signal. Bottom: the square root of energy fluence.}
\label{fig:e_field_rec_hist}
\end{figure}

We define the signal to noise ratio as SNR = $S_{\mathrm{peak}}$/RMS$_{\mathrm{noise}}$, where $S_{\mathrm{peak}}$ is the maximum  of the Hilbert envelope of the voltage, and RMS$_{\mathrm{noise}}$ is the root mean square of the noise, calculated over a 250 ns noise window located 500 ns after the signal maximum. We select the antennas having an SNR greater than 5 in any polarization and whose signal peak falls within the expected signal arrival time. Additionally, to ensure reliable direction and energy reconstruction within the pipeline, we require each event to have at least 5 antennas satisfying these criteria. Applying this selection results in 54\% of the simulated events being retained. 

Figure \ref{fig:e_field_rec} illustrates an example of electric field reconstruction using the analytical least-squares method. Fluctuations at certain frequencies are observed due to the small antenna response at those frequencies. However, when converting back to the time domain, these fluctuations have minimal impact on the trace. The reconstructed trace closely overlaps with the simulated trace, demonstrating that the precision and accuracy of the direction reconstruction is not affected by the electric field reconstruction.

To provide a quantitative assessment of the performance of this method to reconstruct the electric field, we applied it to all events in the air shower library described in Section \ref{sim_set}. In the top panel of Figure \ref{fig:e_field_rec_hist}, the maximum of the envelope ($E_{\mathrm{peak}}$) of the reconstructed total electric field is compared with the simulated one. The zero median and a mean of 0.00 suggest that this reconstruction is unbiased. The 68\% confidence interval with a range of [-0.02, 0.02] and a small standard deviation of 0.04 further highlight the high accuracy and minimal variation of the reconstruction.

Energy fluence quantifies the energy deposited per unit area on each antenna within the signal window. The total measured radiation energy is obtained by spatially integrating over the entire array within its operational frequency band. This quantity is fundamental to energy reconstruction using radio signals \cite{Glaser:2016}.
It can be expressed in terms of the reconstructed electric field as

\begin{equation}
\Phi = c \cdot \epsilon_0 \cdot  \int \vec{\boldsymbol{\mathcal{E}}}^2(t) dt 
\label{eq:fluence}
\end{equation}
where c the speed of light in vacuum, $\epsilon_0$ the permittivity in vacuum and t is the time. For this analysis, we selected a 100 ns time window centered at the signal maximum and a 100 ns window at the end of the trace to estimate the fluence of the background noise. To determine the energy fluence, the noise fluence is subtracted from the fluence calculated for the signal window. The distribution of the relative error of the energy fluence for the reconstructed electric field, shown at the bottom of Figure \ref{fig:e_field_rec_hist}, an estimate of the minimum uncertainty we would get in the energy reconstruction using this electric field reconstruction method. 

The relative error distribution of the energy fluence reveals a slight bias in the energy fluence calculation, with a mean of 0.02 and a median of 0.00. The 68\% confidence interval range from -0.04 to 0.01, and the  standard deviation is 0.06, both of which are larger compared to that of the electric field maximum amplitude. This discrepancy is attributed  to the accumulation of errors during the integration of the electric field over time.



\section{Direction reconstruction}

Achieving a good angular resolution is essential to trace the origins of cosmic rays. To reconstruct the primary particle incoming direction, our analysis relies on two key parameters from the detected signals: the peak time, which indicates the arrival time, and the signal strength. The relative differences in peak times across the triggered antennas allow for the reconstruction of the apparent source of the radio emission, closely related to the position of the shower maximum $X_\mathrm{max}$. By incorporating the signal strength, the complete wavefront pattern at ground level can be obtained \cite{Decoenethesis:2020,decoene2021reconstruction, GlaserA:2019, Felix:2023}. From this, the shower axis can be deduced, which inherently includes the shower core position—crucial for reconstructing both the arrival direction and the primary energy (E$_\mathrm{pri}$), as discussed in the next chapter.

Previous studies \cite{corstanje2015shape} identified the hyperbola as an effective model for the wavefront shape of EAS. 
For inclined air showers, it has been shown that the wavefront can be modeled as spherical, given that the emission region is relatively small compared to the total propagation distance \cite{decoene2023radio}. 

To address the complexities of inclined air showers, we apply a methodology developed in \cite{Decoenethesis:2020,decoene2021reconstruction}, specifically designed to achieve high precision in such scenarios. This method begins with determining the apparent position of the emission source $R_{X_{\mathrm{max}}}= (x_{X_\text {max}}, y_{X_\text {max}},$ $ z_{X_\text{max}})$, using the spherical wavefront model. Subsequently, the shower-axis is derived by utilizing the emission pattern, defined by an electric field angular distribution function (Equation \ref{eq:f_adf}), enabling an accurate determination of the shower direction.

For this analysis, both the peak value and peak time of the signal are required. The peak time is determined from the voltage trace with noise, while the peak amplitude is obtained from the reconstructed electric field, as it would be done in real experimental conditions.

\subsection{Spherical wavefront model}
The radio emission can be approximated as coming from a point source in the vicinity of the shower maximum $X_\mathrm{max}$.  In the spherical wavefront model to determine position of the emission source $R_{X_{\mathrm{max}}}$, the expected arrival time at a given antenna can be expressed as: 

\begin{equation}
\frac{c}{n_i}\left(t_i-\bar{t}\right)= \sqrt{\left(x_i-x_{_\text {Xmax }}\right)^2+\left(y_i-y_{_\text {Xmax }}\right)^2+\left(z_i-z_{_\text {Xmax}}\right)^2}
\label{eq:swf}
\end{equation}
where c is speed of light, $n_i$ is the average index of refraction between emission source $R_{X_\mathrm{max}}$ and the antenna and $x_i$, $y_i$, $z_i$ and $t_i$ are its corresponding coordinates and signal arrival time. The parameter $\bar{t}$, is the average peak time, and is derived from the known arrival times. The minimiation of $\chi^2=\frac{c}{n_i}\left(t_i-\bar{t}\right) - \sqrt{\left(x_i-x_{_\text {Xmax }}\right)^2+\left(y_i-y_{_\text {Xmax }}\right)^2+\left(z_i-z_{_\text {Xmax}}\right)^2}$ gives the best guess for $R_{X_{\mathrm{max}}}$.


\subsection{The Angular Distribution Function}
\label{sec:ADF}

The angular distribution function (ADF) of the electric field amplitude within the footprint comprises two distinct components. The first component, denoted as $f^{ch}$, describes the inherent Cherenkov ring structure in the emission pattern. The second component, labeled as $f^{asym}$, accounts for the asymmetry introduced by the superposition of geomagnetic and charge-excess effects, each characterized by unique polarization patterns \cite{Decoenethesis:2020}.

This can be represented as following :

\begin{equation}
f^{ADF}\left(\omega_i, \phi_i, l_i, \alpha , \delta \omega, A_1 \right) = \frac{A_1}{l_i} f^{ch}\left(\omega_i, \delta \omega\right) f^{asym}\left(\phi_i, \alpha\right)
\label{eq:f_adf}
\end{equation}

with

\begin{equation}
\begin{aligned}
        f^{ch}\left(\omega_i, \delta \omega \right) = \frac{1}{1+4\left[\frac{\left(\tan \omega_i / \tan \omega_c\right)^2-1}{\delta \omega}\right]^2}\\ \label{eq:f_ch}
    f^{asym}\left(\phi_i, \alpha\right) = 1 + \mathcal{B} \sin^2 \alpha \cos \phi_i
\end{aligned}
\end{equation}

where $\omega_i$ represents the opening angle between the shower axis and the antenna, measured from the emission point. Here, $\omega_c$ denotes the Cherenkov angle, corresponding to the angle of maximum signal, and $\delta \omega$ is a free parameter related to the width of the Cherenkov ring. $\mathcal{B}$ denotes the geomagnetic asymmetry strength (set to 0.01), $\phi_i$ is the polar angle of the antenna position in the shower plane, and $\alpha$ is the angle between the shower direction and the magnetic field. Finally, $A_1$ is the parameter related to the amplitude of the signal, $l_i$ represents the distance from the antenna to emission source $R_{X_{\mathrm{max}}}$, which is associated with the attenuation due to propagation of the spherical wavefront. As a result, antennas located closer to $R_{X_{\mathrm{max}}}$ receive stronger signals than the ones farther away, which allows the ADF model to include the early-late effect \cite{Felix:2023}.

Given the emission source $R_{X_{\mathrm{max}}}$, the quantities $\omega_i$, $\phi_i$, and $\alpha$ are all derived from air shower direction $\theta$ and $\varphi$. Therefore, when fitting this function, the shower direction, and amplitude factor A can be accurately determined. 

\subsection{Selection criteria}
In addition to the selection criteria discussed in the previous section for electric field reconstruction, and a cut on the goodness of the $\chi^2$ fit, we introduce two additional cuts which will be needed later for energy reconstruction. Since the energy fluence fit relies on both the direction and emission source $R_{X_\mathrm{max}}$, the following cuts are applied together to ensure a good reconstruction:

\begin{itemize}
\item The altitude of the shower core, calculated from the reconstructed direction from a plane wave reconstruction and the reconstructed emission source $R_{X_\mathrm{max}}$, is constrained to be between 500 m and 2000 m, assuring a reconstruction consistent with the fact that the ground altitude is 1068 m. 
\item The fitted amplitude of the energy fluence must be less than $10^9$, a threshold determined from the Monte Carlo truth for this library.
\end{itemize}

As a result, 77\% of the traces in selected events pass these quality cuts.

\subsection{$R_{X_{\mathrm{max}}}$ resolution}

\begin{figure*}[htbp]
    \centering
    \begin{minipage}{0.45\textwidth}
        \centering
        \includegraphics[width=\textwidth]{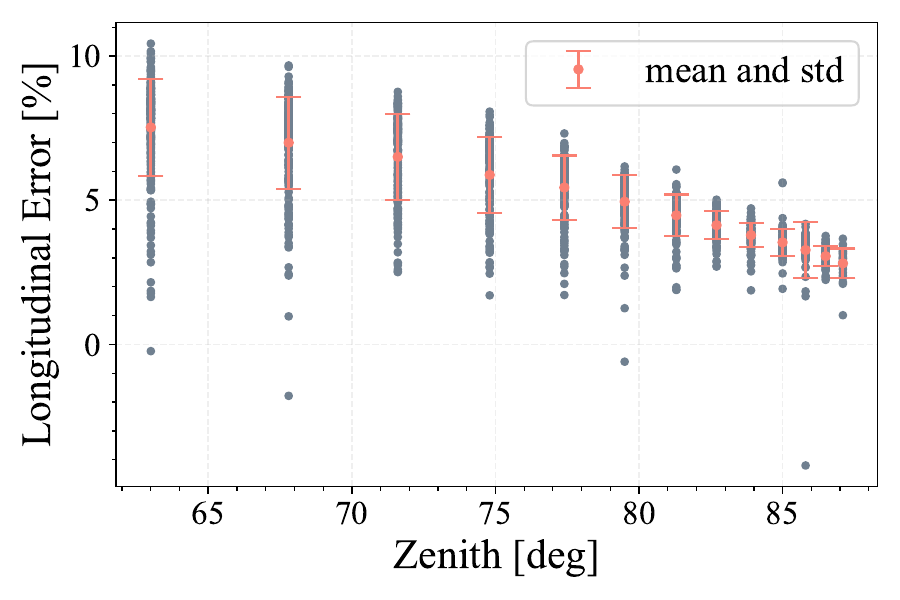}
    \end{minipage}%
    \hspace{0.05\textwidth}
    \begin{minipage}{0.45\textwidth}
        \centering
        \includegraphics[width=\linewidth]{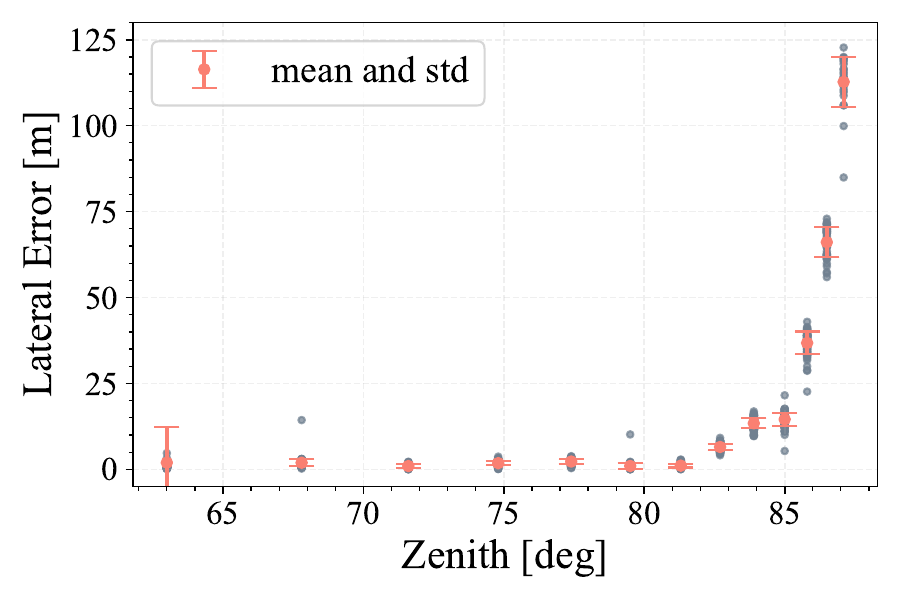}
    \end{minipage}
    \caption{Left: Relative longitudinal error in the $R_{X_{\mathrm{max}}}$ determination as function of zenith angle. Right: lateral error as a function of zenith angle.}
    \label{fig:long_lat_err}
\end{figure*}

\begin{table*}[htbp]
\centering
\caption{Zenith-dependent lateral distribution statistics}
\begin{tabular}{|l|c c c c c c c c c c c c c|}
\hline
Zenith [$^\circ$]     & 63.00 & 67.80 & 71.60 & 74.80 & 77.40 & 79.50 & 81.30 & 82.70 & 83.90 & 85.00 & 85.80 & 86.50 & 87.10 \\
\hline
Longitudinal Mean [\%]             & 7.53  & 7.00  & 6.51  & 5.88  & 5.44  & 4.95  & 4.48  & 4.14  & 3.79  & 3.54  & 3.27  & 3.06  & 2.81 \\
Longitudinal Std [\%]       & 1.68  & 1.60  & 1.49  & 1.31  & 1.11  & 0.92  & 0.72  & 0.48  & 0.42  & 0.47  & 0.98  & 0.34  & 0.51 \\

Lateral Mean [m]            & 1.96  & 1.95  & 0.98  & 1.82  & 2.28  & 1.00  & 1.09  & 6.52  & 13.40 & 14.56 & 36.78 & 66.09 & 112.72 \\
Lateral Std  [m]             & 10.25 & 1.02  & 0.48  & 0.65  & 0.69  & 0.90  & 0.55  & 0.91  & 1.49  & 1.95  & 3.34  & 4.32  & 7.36 \\
\hline
\end{tabular}
\label{tab:zenith_rows}
\end{table*}

We used the iminuit package in Python to perform the minimization for fitting in direction reconstruction \cite{iminuit}.

The longitudinal error of $R_{X_{\mathrm{max}}}$ represents the discrepancy along the shower’s trajectory, while the lateral error quantifies the difference perpendicular to this direction. For very inclined showers, the longitudinal development can extend over a much larger distance than the lateral distribution does, hence, the lateral error is significantly smaller than the longitudinal error. Therefore, we present the relative error for the longitudinal component and the absolute error for the lateral component.

The left panel of Figure \ref{fig:long_lat_err} displays the relative longitudinal error as a function of zenith angle, following the conventional cosmic ray direction, where the zenith angle ranges from 0 to 90$^\circ$. A positive bias is observed in the longitudinal estimation, which gradually diminishes with increasing zenith angle. Corresponding values are summarized in Table \ref{tab:zenith_rows}. This indicates that the reconstructed $R_{X_\mathrm{max}}$ is larger than the true value, meaning that the reconstructed position is farther from the observer than the true position.
The decreasing offset with increasing zenith angle is attributed to the fact that, for inclined showers, the magnitude of $R_{X_\mathrm{max}}$ becomes significantly larger, leading to smaller relative errors. This trend is also reflected in the corresponding standard deviation values presented in Table \ref{tab:zenith_rows}.

The right panel of Figure \ref{fig:long_lat_err} shows the lateral error as a function of zenith angle. For zenith angles above 80$^\circ$, a growing deviation is observed. This is due to the increased computational cost of determining the effective index of refraction individually for each antenna in highly inclined air showers. As a practical compromise, an average effective index is used, which introduces noticeable deviations at large zenith angles. Nevertheless, the standard deviation remains moderate—around 10 meters—across the full zenith range.

This demonstrates that, overall, our analytical $\chi^2$ minimization method enables accurate electric field reconstruction, allowing the longitudinal distance to be determined using the spherical wavefront model with an average relative error with standard deviation of 0.9\%. The lateral deviation remains around 2.6 meters on average. These results are consistent with the findings of \cite{Decoenethesis:2020,decoene2021reconstruction,decoene2023radio}, which used simulated electric fields.





\subsection{Angular resolution}

\begin{figure}[htbp]
\centering
\includegraphics[width=0.45\textwidth]{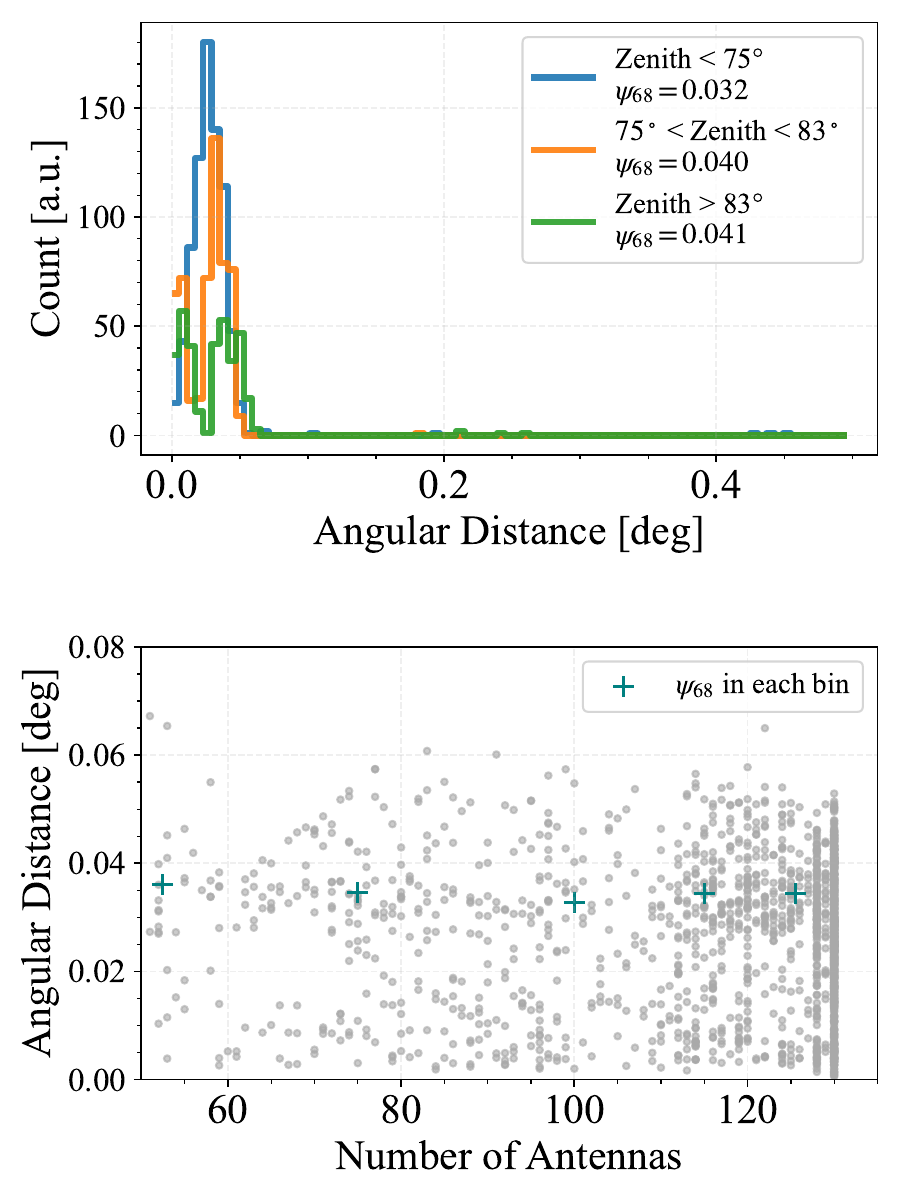}
\caption{Top: Angular resolution distribution. For most of reconstructed events, the angular resolution is less than 0.1$^\circ$. The colors represents the range of zenith angle. Bottom: Angular resolution w.r.t. the number of triggered antennas. }
 \label{fig:ang_rec}
\end{figure}

The emission source $R_{X_\mathrm{max}}$ is used as input for the ADF minimization to get the reconstructed zenith angle ($\theta_\mathrm{rec}$) and azimuth angle ($\varphi_\mathrm{rec}$) as outlined in section \ref{sec:ADF}. The evaluation of the reconstruction quality is based on the angular distance, defined as

\begin{equation}
    \cos\psi=\cos\theta_\mathrm{rec}\cos\theta_\mathrm{sim}
    + \cos(\varphi_\mathrm{rec}-\varphi_\mathrm{sim})\sin\theta_\mathrm{sim}\sin\theta_\mathrm{rec}
\label{eq:angular dis}
\end{equation}

where $\theta_\mathrm{sim}$ and $\varphi_\mathrm{rec}$ represent true zenith and azimuth angles.

Figure \ref{fig:ang_rec} presents the distribution of angular resolution. In the top panel, we observe a variation in resolution with respect to the zenith angle, although most of the resolutions remain below 0.04$^\circ$. For showers with the same energy, the signal strength is lower for inclined showers, resulting in fewer triggered events for zenith angles greater than 83$^\circ$. The bottom panel shows the resolution as a function of the number of triggered antennas. Given the star-shaped layout, which covers the entire emission region’s footprint, we observe only a few events with less than 50 triggered antennas. For cases with more than 50 triggered antennas, since the shower footprint is sufficiently sampled, accurate direction reconstruction is always feasible, demonstrating that the precision of direction reconstruction can be achieved through electric field reconstruction.


\section{Energy reconstruction}
\begin{figure}[htbp]
\centering
\includegraphics[width=0.45\textwidth]{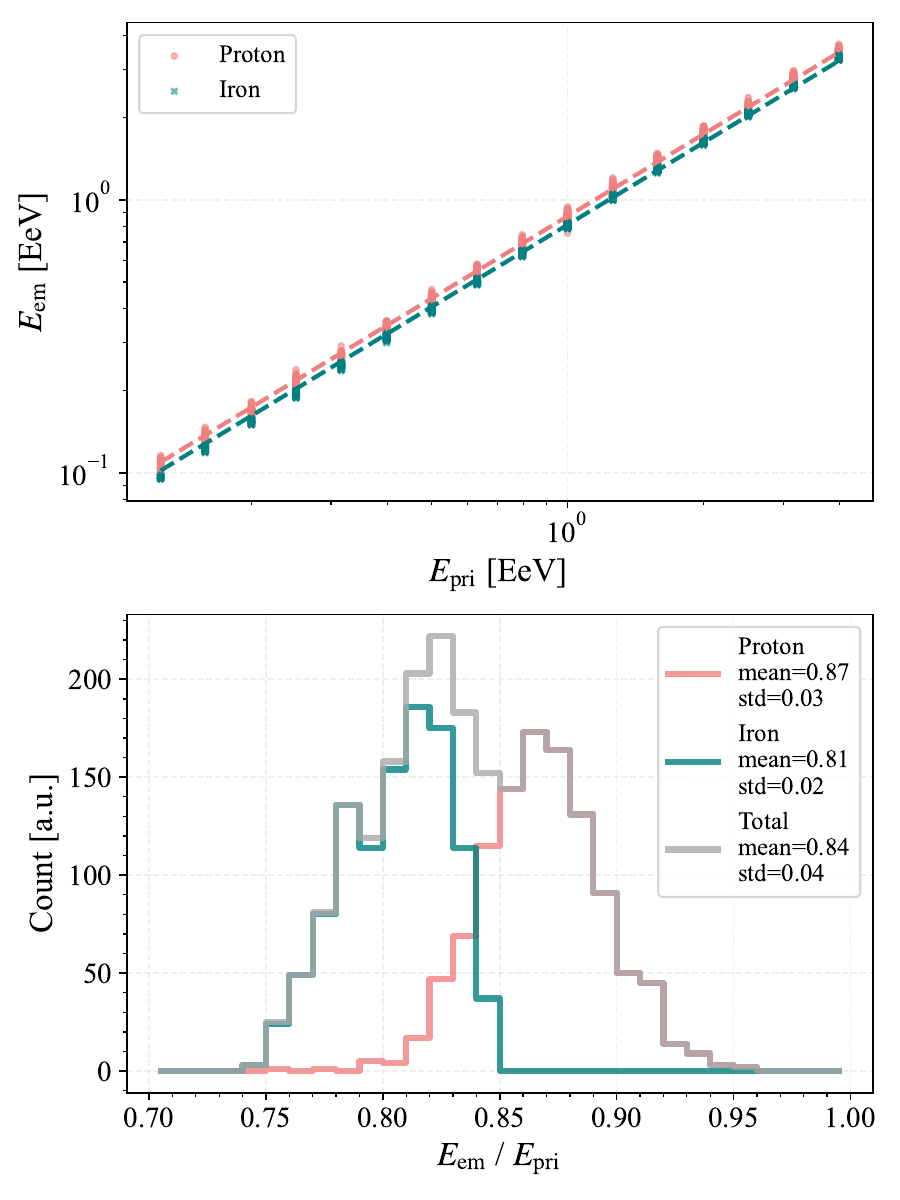}
\caption{Top panel: Relationship between the primary particle energy E$_\mathrm{pri}$ and the electromagnetic energy $E_\mathrm{em}$ induced by the EAS. Bottom panel: Histogram of the ratio between the electromagnetic energy $E_\mathrm{em}$ and the primary energy $E_\mathrm{pri}$.  }  
\label{fig:E_ratio}
\end{figure}

Many methods have been investigated to reconstruct the electromagnetic energy of air showers. 
Initial efforts to parameterize the one-dimensional lateral distribution of the radio signal~\cite{Allan1971} were implemented by experiments such as CODALEMA and LOPES, which used simple, narrow-band antennas to detect vertical air showers~\cite{LOPES:2005ipv, Ardouin:2005qe}. In modern experiments such as LOFAR, with more antennas deployed, a better determination of the signal profile on the ground could be made using a 2D lateral distribution function (LDF) \cite{Anna:2015}.

Radio emission experiences little attenuation in the atmosphere and carries the complete information about the  air shower development.  Radio antennas measure the electromagnetic component of air showers, which represents most of the primary energy $E_\mathrm{pri}$. For the air shower library used in this work, the electromagnetic energy ($E_\mathrm{em}$) is determined by summing up the energy of the low energy particles discarded from the simulation and the longitudinal energy deposits for gamma rays, electrons, and positrons.
The top panel of Figure \ref{fig:E_ratio} illustrates the relationship between  $E_\mathrm{pri}$  and  $E_\mathrm{em}$ for the extreme cases of protons and irons. The bottom panel show the ratio of $E_\mathrm{em}$/E$_\mathrm{pri}$ for proton and iron primaries. The electro-magnetic component constitutes, on average, 87\% of the primary energy in the proton air showers, and 81\% in the Iron air showers. The overall deviation remains small, with a standard deviation of 0.04 when all  events are considered together, further reconfirming the small fluctuations found for a given $E_\mathrm{pri}$ in the top panel. The 84\% overall proportion is close to the study shown in \cite{Mulrey:2020oqe}, whichreported an 85\% overall proportion in the LOFAR energy range, also showing a dependence on the primary particle type. 


With the development of tools like CoREAS \cite{Coreas2013} and ZHAireS \cite{ZHAireS2012} in radio simulation, it was found that air density at the shower maximum must be taken into account as a second-order dependence to further correct the energy estimation \cite{Glaser:2016}, that we will present in section \ref{sec:energycorrections}. Furthermore, asymmetries produced by the emission mechanisms and propagation were clearly observed in simulation and verified on real data \cite{nelles2015radio}. Methods were initially developed in AERA and then refined in LOFAR and TUNKA to study the asymmetries caused by the  decomposition of the electric field into transverse current and charge-excess contributions \cite{PierreAuger:2014ldh, Schellart:2014oaa, Kostunin:2016}. These methods can be used to determine the radiation energy from both effects \cite{Glaser:2016}, by parameterizing the charge excess fraction and correcting the geomagnetic radiation energy for geometry in the estimation of E$_\mathrm{em}$. This methodology has been  widely  employed in LOFAR \cite{Mulrey:2020oqe}. At the Pierre Auger Observatory, the charge-excess fraction has been reformulated as a correction to obtain a rotational symmetric LDF. With this parameterization, the geomagnetic radiation fluence is determined and applied to correct the geomagnetic radiation energy \cite{Felix:2023}. In this work, we follow the logic of this approach, as will be described in the next section.

\subsection{Air density and magnetic corrections}
\label{sec:energycorrections}
The overall geomagnetic radiation energy $E_\mathrm{geo}$ is calculated by integrating the geomagnetic radio energy fluence ($\Phi_\mathrm{geo}$) over the radio footprint on the ground, 
\begin{equation}
E_\mathrm{geo} = 2\pi \cdot  \int ^{\infty} _{0}\Phi_\mathrm{geo}(r) rdr
\end{equation}
where $\Phi_\mathrm{geo}$ is the geomagnetic component of the total energy fluence calculated with Equation \ref{eq:fluence} on the geomagnetic component of the reconstructed electric field and parametrized as a function of the distance to the shower core \cite{Glaser:2016}. The factor $d_{X_\mathrm{max}}$ is the distance from shower core to emission point $R_{X_\mathrm{max}}$, and is used to get $E_{geo}$ normalized to the emission point, in order to make it comparable among different showers. 

With the Cherenkov term of  ADF description ($f^{ch}$), it is expressed as
\begin{equation}
     E_\mathrm{geo}=d_\mathrm{X_{max}}^2\int A_2 \cdot f^{ch}(\omega)d\omega
\end{equation}

were A$_2$ is the amplitude of the energy fluence obtained with the ADF description of footprint. 

\begin{figure}[htbp]
\centering
\includegraphics[width=0.5\textwidth]{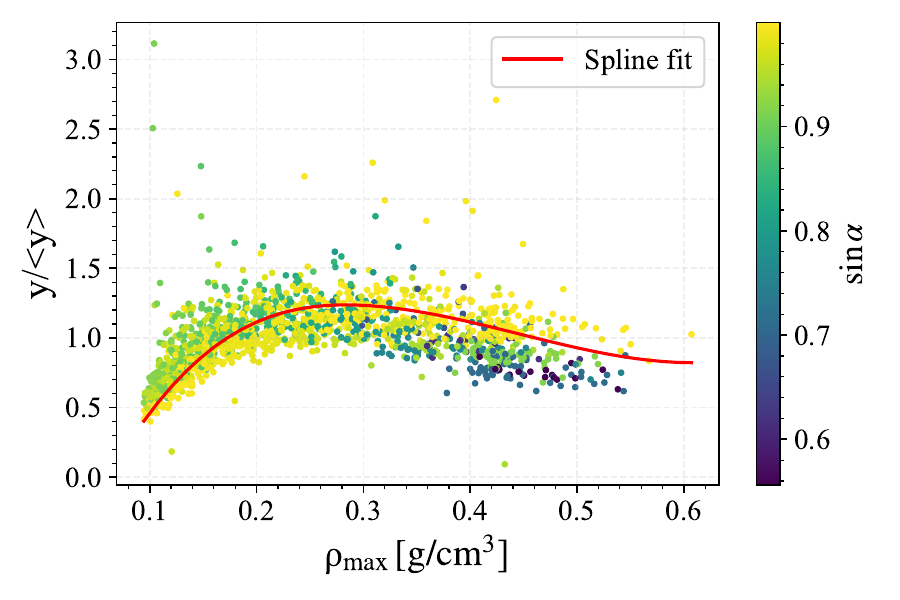}
\caption{Air density correction for geomagnetic energy fluence. The air density is evaluated at the reconstructed shower maximum $\mathrm{X_{max}}$. And $\mathrm{y}=E_{\mathrm{geo}}/( (E_{\mathrm{em}}/\mathrm{EeV})^2 \cdot \sin\alpha)$, where $\mathrm{<y>}$ denotes the average value across all simulations used for normalization. The red line represents the spline fit to the simulation data under ideal conditions, accounting for the full footprint and direction.}
\label{fig:air_den}
\end{figure}

\begin{figure*}[hbtp]
    \centering
    \begin{minipage}{0.48\textwidth}
        \centering
        \includegraphics[width=\textwidth]{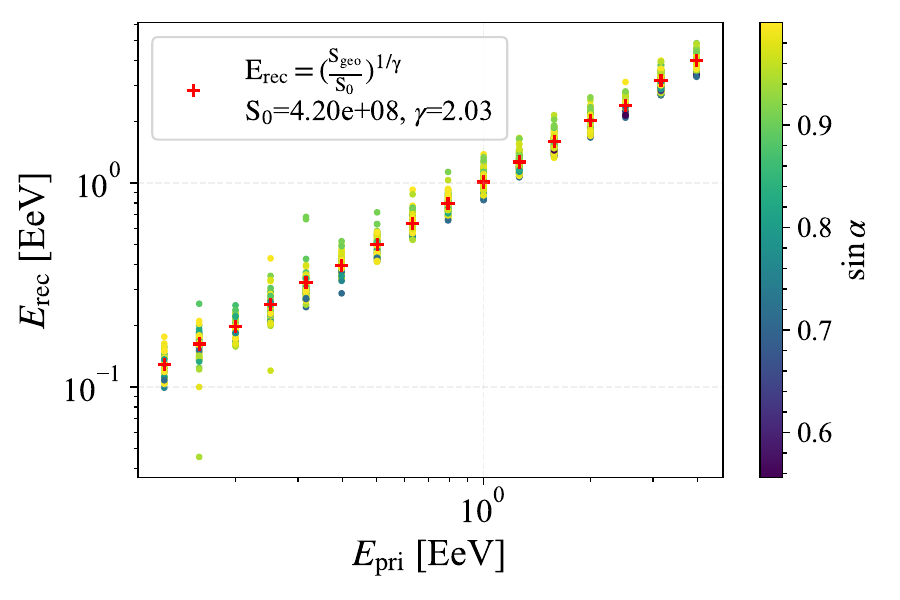}
    \end{minipage}%
    \hspace{0.05\textwidth}
    \begin{minipage}{0.40\textwidth}
        \centering
        \includegraphics[width=\linewidth]{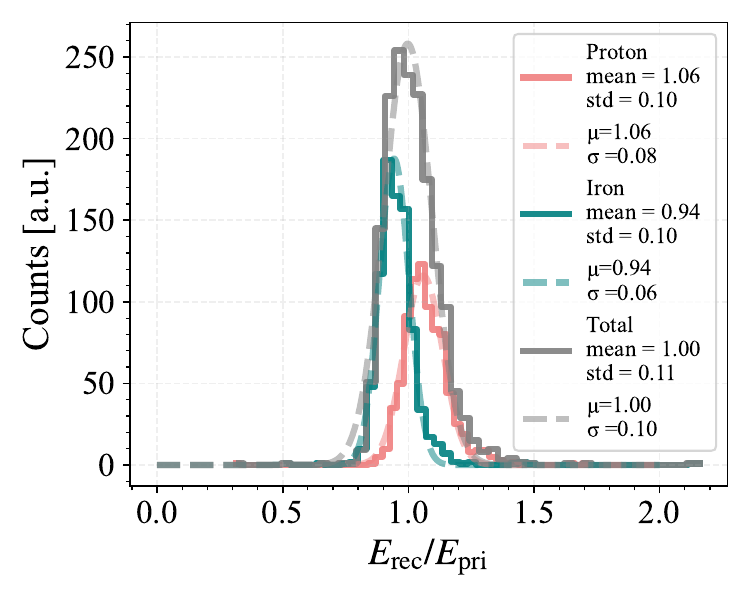}
    \end{minipage}
    \caption{Reconstruction of primary energy from radio emission of EAS. Left panel: Scatter plot of reconstructed primary energy as a function of true primary energy, with color indicating the value of sin$\alpha$ for each simulation.  
    The red cross denotes the mean reconstructed energy for each primary energy bin. Right panel: Histogram of the ratio of reconstructed to true primary energy (solid line), with the dashed line representing a Gaussian fit.} 
    \label{fig:energy_rec}
\end{figure*}

The geomagnetic effect is the main radio emission in air showers, and is produced by the deflection of charged particles in the local geomagnetic field of the experiment. Its magnitude is proportional to the Lorentz force, hence the absolute value of B$\cdot \mathrm{sin}\alpha$, where B is the magnetic field strength, $\alpha$ is the angle between the magnetic field and the direction of the shower.  Therefore, the coherent geomagnetic radiation energy is proportional to B$^2\mathrm{sin}^2\alpha$. The relatively high geomagnetic field strength present in the site for which the simulation were made, would made us expect a larger coherent emission at large zenith angles. However, a competing effect appears, resulting in the relative drop at low densities seen in Figure \ref{fig:air_den}. The electrons and positrons start radiating from the moment they are generated, and loose energy mainly due to multiple elastic scatterings in the air until being absorbed.  Their attenuation length in air is $l_\mathrm{rad} = 36.7$ g cm$^{-2}/\varrho_\mathrm{air}$.
If the air density is high, the electrons and positrons are absorbed in a short distance, as is the case in vertical showers, where the maximum of the cascade occurs deep in the atmosphere. In inclined showers, the maximum of the cascades occurs at much higher altitude, with lower air density, so the attenuation length increases significantly. When the magnetic force is high enough, electrons and positrons can gyrate along much longer arcs, producing an extended lateral particle distribution and radiating energy for a  longer time. Since the polarization of this radiation is no longer linear, a loss of coherence is generated, resulting in a new signal pattern in the so called geosynchrotron regime.  This particular configuration has not been thoroughly examined in the context of energy reconstruction before the detailed discussions in \cite{James:2022, geosynchrotron2024}, revealing a notable drop in the relative strength of the geomagnetic emission at low air density, particularly corresponding to very large zenith angles.  



Since the different shower geometries will produce different radiation amplitudes depending on the orientation with respect to the local magnetic field and air density, the geomagnetic radiation energy needs to be corrected before it can be correlated with the energy of the electromagnetic cascade. 



The corrected geomagnetic radiation energy in this case is computed as:

\begin{equation}
    S_\mathrm{geo}=\frac{E_\mathrm{geo}}{\sin\alpha}\cdot\frac{1}{f(\rho_\mathrm{max})}
    \label{eq:s_geo}
\end{equation}

where we use a spline fit $f(\rho_\mathrm{max})$ to describe the density effect, based on the atmospheric density value at the reconstructed $R_{X_{\mathrm{max}}}$ position, depicted in Figure \ref{fig:air_den}.  With the $\sin \alpha$ term, the correction factor fits well the data. For large $\sin \alpha$, the correction factor drops at low air density, where inclined air showers achieve the shower maximum. The maximal scaling factor is obtained at a critical air density around 0.3 $\mathrm{g/cm^3}$. This correction works in the geosynchrotron regime of inclined air shower, and it differs from the correction for vertical air showers \cite{Felix:2023, geosynchrotron2024}.

\subsection{Estimation of primary energy}

 Once the corrected geomagnetic radiation energy $S_\mathrm{geo}$ is obtained, the electromagnetic energy $E_\mathrm{em}$ can be calculated via a power-law correlation function 
 as shown in \cite{Felix:2023}. This method for the estimation of the primary electromagnetic energy has it's origin on the calibration function in the energy reconstruction work in LOFAR \cite{Mulrey:2020oqe} and on the work on radio detection done by the Pierre Auger Collaboration \cite{maris:2013, PierreAuger:2015hbf, Dembi:2016}.

 The stable proportionality between $E_\mathrm{em}$ and $E_\mathrm{pri}$—with only small composition-dependent fluctuations—allows us to bypass an explicit $E_\mathrm{em}$ estimation step. Instead, the reconstruction is performed directly with respect to $E_\mathrm{pri}$, enabling a simplified and more applicable reconstruction pipeline for autonomous, radio-only experiments.
 

We establish a power law relation for the reconstructed primary energy as follows:
\begin{equation}
    E_\mathrm{rec}=\left(\frac{S_\mathrm{geo}}{S_0}\right)^{1/{\gamma}}
    \label{eq:E_rec_em}
\end{equation}
where $S_0$ is the normalization parameter and $\gamma$ is the spectral index.

In Figure \ref{fig:E_ratio}, the mean ratio $E_{\mathrm{em}}/E_{\mathrm{pri}}$ differs between primary particle types, introducing a bias in the reconstruction of primary energy. But if the primary is indistinguishable, no bias is produced on average (grey line).




\begin{figure*}[ht!]
    \centering
    \begin{minipage}{0.42\textwidth}
        \centering
        \includegraphics[width=\textwidth]{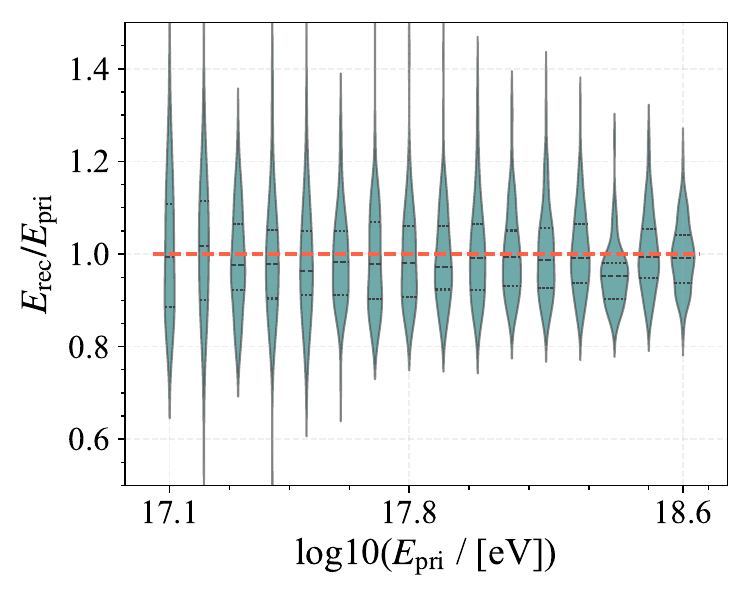}
    \end{minipage}%
    \hspace{0.05\textwidth}
    \begin{minipage}{0.43\textwidth}
        \centering
        \raisebox{4mm}{ 
            \includegraphics[width=\linewidth]{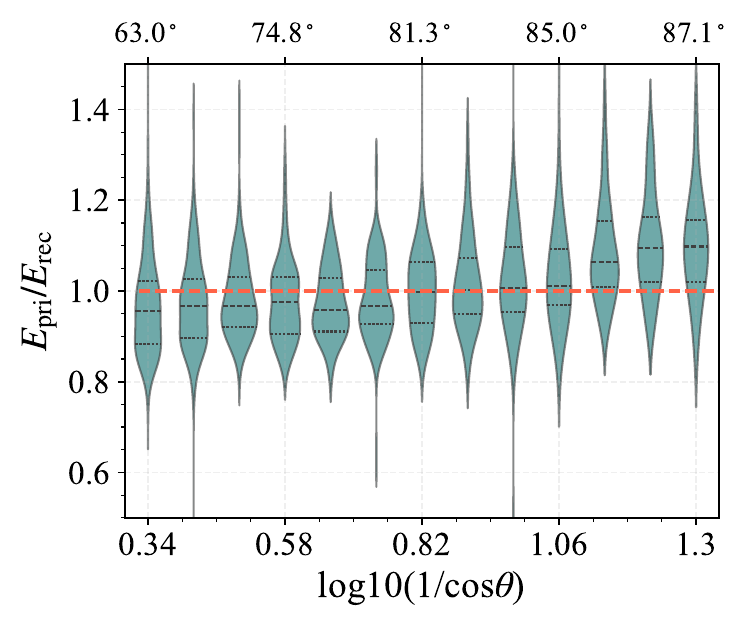}
        }
    \end{minipage}
    \vspace{0.05\textwidth}
    \begin{minipage}{0.42\textwidth}
        \centering
        \includegraphics[width=\textwidth]{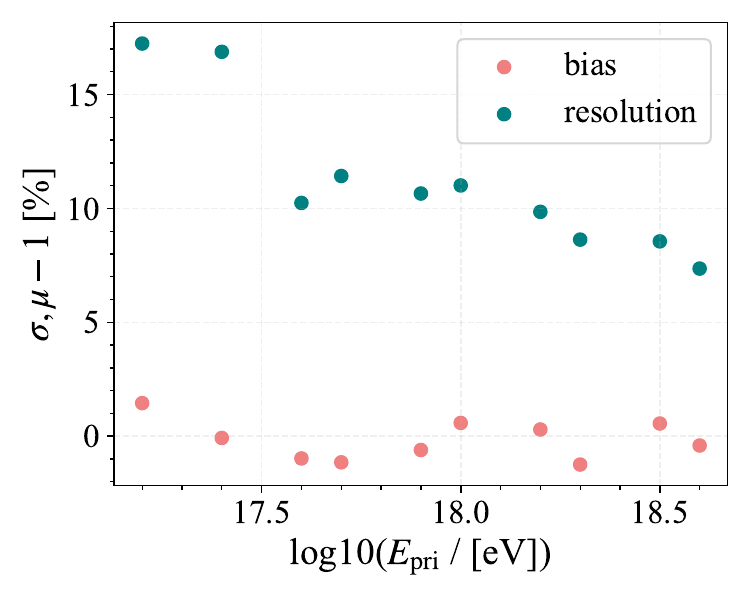}
    \end{minipage}%
    \hspace{0.05\textwidth}
    \begin{minipage}{0.42\textwidth}
        \centering
        \includegraphics[width=\linewidth]{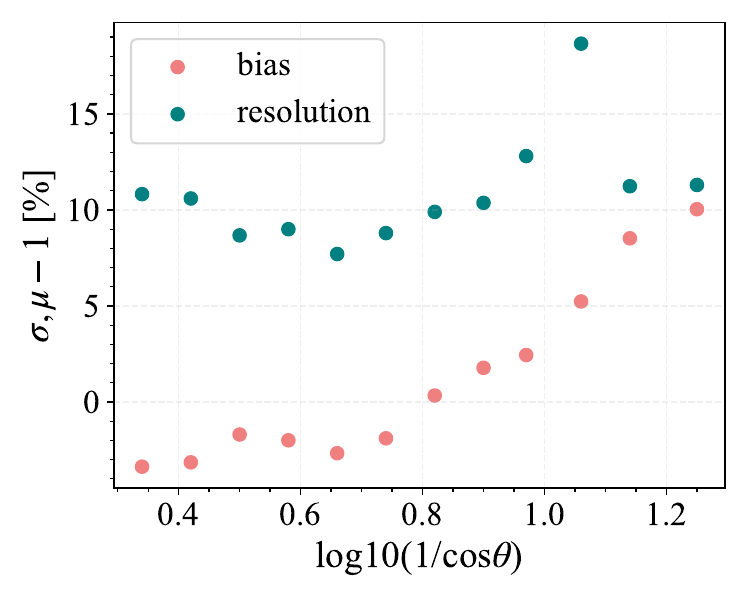}
    \end{minipage}
     
    \caption{Bottom panels: Bias and resolution of the reconstructed electromagnetic energy as a function of true energy (left) and zenith angle (right). Top panels: Corresponding full distributions.}
    \label{fig:energy_zenith}
\end{figure*}

\subsection{Energy resolution}
Given the $E_\mathrm{geo}$ obtained with the ADF method, the $S_\mathrm{geo}$ is derived from Equation \ref{eq:s_geo}. The parameters $S_0$ and $\gamma$ are obtained by minimizing the residual between the reconstructed energy $E_\mathrm{rec}$ and the true primary energy $E_\mathrm{pri}$.  The left panel of Figure \ref{fig:energy_rec} shows the comparison between the reconstructed energy $E_\mathrm{rec}$ and the true primary energy $E_\mathrm{pri}$, with the color scale indicating the value of $\sin \alpha$. With the air density correction applied to $E_\mathrm{geo}$, no significant dependence on $\sin \alpha$ is observed.

To investigate the dependence on primary energy and zenith angle, the full distribution of $E_\mathrm{rec} / E_\mathrm{pri}$ is shown in the top panels of Figure \ref{fig:energy_zenith}, after merging the first four energy bins into two and the last two zenith bins to reduce large variations in event statistics across bins. The resolution $\sigma$ and bias $\mu$ are shown on the bottom panels. This method achieves a resolution better than 10\% in most of the energy range, with some degradation at lower energy. For highly inclined air showers with zenith  angles $\theta>80^{\circ}$, a larger bias is found.  This is consistent with the decreasing performance of the analytical least square method of the electric field in these conditions \cite{Zhang:2025qef}. 

In the right panel, the overall distribution is fitted with a Gaussian. While the total sample shows no significant bias in the reconstructed primary energy, a composition-dependent trend is observed: proton-induced showers tend to be slightly overestimated, and iron-induced showers slightly underestimated. This bias reflects the intrinsic difference in the $E_{\mathrm{em}}/E_{\mathrm{pri}}$ ratio for different primaries. Nevertheless, the overall energy resolution remains at the 10\% level, confirming the robustness of the reconstruction method.

\begin{figure}[ht!]
\centering
\includegraphics[width=0.50\textwidth]{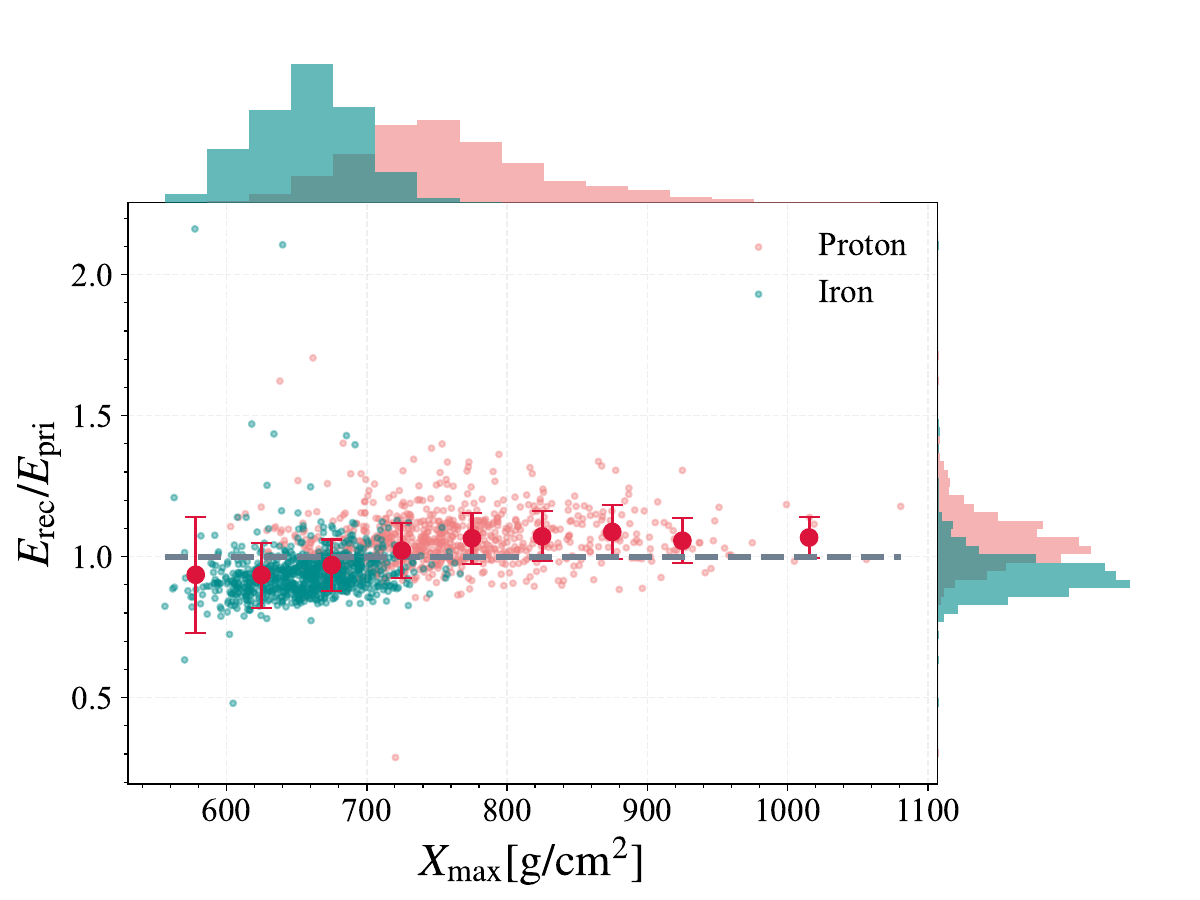}
\caption{$E_{\mathrm{em}}/E_{\mathrm{pri}}$ as function of $\mathrm{X_{max}}$. The dots and error bars indicate the mean and standard deviation in each  $X\mathrm{_{max}}$ bin, revealing an $X\mathrm{_{max}}$-dependent bias.  The top and right histograms display respectively the distributions of $X\mathrm{_{max}}$ and $E_{\mathrm{em}}/E_{\mathrm{pri}}$.}
\label{fig:Erec_Xmax}
\end{figure}

Given the composition-dependent bias in energy reconstruction and the strong correlation between $X_{\mathrm{max}}$ and the primary particle type, a corresponding bias in $X_{\mathrm{max}}$ is also observed, as shown in Figure~\ref{fig:Erec_Xmax}. Nevertheless, the reconstruction resolution remains consistent across different $X_{\mathrm{max}}$ values.

\section{Conclusions and outlook}

We present a comprehensive study of radio detection reconstruction for UHECRs, first reconstructing the electric field and then using it to determine the arrival direction and energy of air showers.

We have employed a new analytical $\chi^2$ minimization method based on three-polarization antenna in electric field reconstruction. This approach is motivated by the growing demand for radio detection of cosmic rays and neutrinos at large zenith angles, where capturing the full vector structure of the radio signal is essential. The method achieves a 5\% resolution in the peak electric field amplitude and an 8\% resolution in the energy fluence derived from the reconstructed electric field trace.



To reconstruct the arrival direction of inclined air showers, we employ a spherical wavefront approximation. Within this framework, the emission source position is first determined from the peak time and then used to fit the radio footprint of the reconstructed electric field peak with the ADF, from which the arrival direction is derived.
We observe a dependence of the angular resolution on the zenith angle, with precision decreasing as the zenith angle increases. Nonetheless, applying a stringent selection criterion—requiring at least five antennas with SNR above 5—the method achieves an overall angular precision of approximately 0.04$^\circ$ across the full zenith angle range.


For energy reconstruction, we extract the geomagnetic component of the radio emission. This component, utilized in the ADF during direction reconstruction, serves as the basis for estimating the radiation energy from the reconstructed electric field. Since our focus is on inclined showers developing in strong magnetic fields—where the radio emission enters into the geosynchrotron regime—we first apply a modified air density correction to the geomagnetic component, scaled by $\sin\alpha$.
After applying an air density correction, the corrected geomagnetic radiation energy is obtained, serving as a parameter in estimating the electromagnetic energy $E_\mathrm{em}$. We have thoroughly investigated the energy resolution's dependence on primary energy and zenith angles. Statistical analyses show that our method achieves an energy resolution better than 10\% across most of the parameter space, with degradation occurring when the primary energy falls below 10$^{17.5}$ eV or when the zenith angle exceeds approximately 85$^\circ$. These conclusions are valid for both the proton and iron air showers.



It's important to note that our analysis relies on simulated signals on a star-shape pattern that guarantees complete and symmetric footprint sampling, which may present an optimistic perspective. Future work will involve a comprehensive analysis of the entire reconstruction process with real data to enhance our understanding.


\noindent

\begin{acknowledgements}
We would like to thank Tim Huege, Bruno L. Lago, Xin Xu, Pengfei Zhang, Xiaoyuan Huang, Jiale Wang and all the other GRAND members for the related discussions.  This work is supported by the National Natural Science Foundation of China (Nos. 12273114), the Project for Young Scientists in Basic Research of Chinese Academy of Sciences (No. YSBR-061), and the Program for Innovative Talents and Entrepreneur in Jiangsu, and High-end Foreign Expert Introduction Program in China (No. G2023061006L). C.Z thanks Zhuo Li and Ruoyu Liu for their support and discussions during this work. 

\end{acknowledgements}


\bibliographystyle{spmpsci}  
\bibliography{biblio.bib}

\bibliography{}

\begin{thebibliography}{9}
\bibitem{Gaisser:2016uoy}
T.~K.~Gaisser, R.~Engel and E.~Resconi, Cosmic Rays and Particle Physics: 2nd Edition, Cambridge University Press, 2016, ISBN 978-0-521-01646-9.

\bibitem{TelescopeArray:2015dcv}
R.~U.~Abbasi \textit{et al.}, [Telescope Array], Astropart. Phys.  \textbf{80}, 131-140 (2016).

\bibitem{PierreAuger:2018pmw}
A.~Aab \textit{et al.} [Pierre Auger],  JCAP \textbf{10}, 026 (2018).

\bibitem{PierreAuger:2014jss}
A.~Aab \textit{et al.} [Pierre Auger], JCAP \textbf{08}, 019 (2014).

\bibitem{Huege:2023pfb}
T.~Huege [Pierre Auger], EPJ Web Conf. \textbf{283}, 06002 (2023).

\bibitem{GRAND:2018iaj}
J.~\'Alvarez-Mu\~niz \textit{et al.} [GRAND], Sci. China Phys. Mech. Astron. \textbf{63}, no.1, 219501 (2020).

\bibitem{Zeolla:2023phg}
A.~Zeolla, J.~Alvarez-Mu\~niz, A.~Cummings, C.~Deaconu, V.~Decoene, K.~Hughes, R.~Krebs, A.~Ludwig, K.~Mulrey and E.~Oberla, \textit{et al.} PoS ICRC2023, 1019 (2023).

\bibitem{Kahn:1966}
F.~D. Kahn and I. Lerche, Proceedings of the Royal Society of London. Series A. Mathematical and Physical Sciences 289 (1966) 206.

\bibitem{Askaryan:1962}
G.~A. Askaryan, Journal of the Physical Society of Japan Supplement 17 (1962) 257.

\bibitem{Aab:2014}
A. Aab et al. [Pierre Auger], Phys. Rev. D \textbf{89}, 052002 (2014).

\bibitem{frank:2016}
F. G. Schröder, Progress in Particle and Nuclear Physics 93 (2017) 1–68. 

\bibitem{Horandel:2019qwu}
J.~R.~H\"orandel, EPJ Web Conf. \textbf{216}, 01003 (2019).


\bibitem{Schluter:2020tdz}
F.~Schl\"uter, M.~Gottowik, T.~Huege and J.~Rautenberg, Eur. Phys. J. C \textbf{80}, no.7, 643 (2020).

\bibitem{James:2022}
C. W. James, Phys. Rev. D \textbf{105}, 023014 (2022).

\bibitem{geosynchrotron2024}
S. Chiche, C. Zhang, F. Schl\"uter, K. Kotera, T. Huege, K.D. de Vries et al., Phys. Rev. Lett. \textbf{132} (2024) 231001.

\bibitem{Schellart:2014oaa}
P.~Schellart, S.~Buitink, A.~Corstanje, J.~E.~Enriquez, H.~Falcke, J.~R.~H\"orandel, M.~Krause, A.~Nelles, J.~P.~Rachen and O.~Scholten, \textit{et al.}, JCAP \textbf{10}, 014 (2014).

\bibitem{Chiche:2022ppi}
S.~Chiche, K.~Kotera, O.~Martineau-Huynh, M.~Tueros and K.~D.~de Vries, Astropart. Phys. \textbf{139}, 102696 (2022).

\bibitem{ZHAireS2012}
J. Alvarez-Muñiz, W.R. Carvalho and E. Zas, Astroparticle Physics \textbf{35} (2012) 325–341.802.

\bibitem{GRANDlib}
R. Alves Batista \textit{et al.}, Comput. Phys. Commun. \textbf{308} (2025), p. 109461.

\bibitem{Glaser:2019}
C. Glaser, A. Nelles, I. Plaisier, C. Welling, S.W. Barwick, D. García-Fernández et al., Eur. Phys. J. C \textbf{79}, (2019) 464. 

\bibitem{Zhang:2025qef}
K.~Zhang, T.~Huege, R.~Koirala, P.~Ma, M.~Tueros, X.~Xu, C.~Zhang, P.~Zhang and Y.~Zhang, 
arXiv:2501.12614 [astro-ph.IM], submitted to JCAP.

\bibitem{Behnke:2013}
O. Behnke, K. Kroninger, G. Schott and T. Schörner-Sadenius,  2013, https://api.semanticscholar. org/CorpusID:60282717.

\bibitem{Lfmap:2007}
E. Polisensky, Memo Series \textbf{111} (2007) 515.

\bibitem{Decoenethesis:2020}
V. Decoene, Ph.D. thesis, Sorbonne universit{\'e}, 2020.

\bibitem{GlaserA:2019}
C. Glaser, S. de Jong, M. Erdmann and J.R. H\"orandel,  Astropart. Phys. \textbf{104} (2019) 64.

\bibitem{Felix:2023}
F. Schl{\"u}ter and T. Huege JCAP \textbf{01}  008 (2023).

\bibitem{corstanje2015shape}
A. Corstanje, P. Schellart, A. Nelles, S. Buitink, J. Enriquez, H. Falcke \textit{et al.},  Astropart. Phys. \textbf{61} (2015) 22.

\bibitem{apel2014wavefront}
W. Apel, J. Arteaga-Velázquez, L. Bähren, \textit{et al.}, JCAP \textbf{09}, 25 (2014).



\bibitem{Kostunin:2016}
D. Kostunin, P. Bezyazeekov, R. Hiller, F. Schröder, V. Lenok and E. Levinson,  Astropart. Phys. \textbf{74} (2016) 79–86.

\bibitem{Mulrey:2020oqe}
K.~Mulrey, S.~Buitink, A.~Corstanje, H.~Falcke, B.~M.~Hare, J.~R.~H\"orandel, T.~Huege, G.~K.~Krampah, P.~Mitra and A.~Nelles, \textit{et al.} JCAP \textbf{11}, 017 (2020).

\bibitem{Allan1971}
H. R. Allan,  Part. Cosmic Ray Phys. \textbf{10} (1971) 169–302.

\bibitem{LOPES:2005ipv}
H.~Falcke \textit{et al.} [LOPES], Nature \textbf{435} (2005), 313-316.

\bibitem{Ardouin:2005qe}
D.~Ardouin, A.~Belletoile, D.~Charrier, R.~Dallier, L.~Denis, P.~Eschstruth, T.~Gousset, F.~Haddad, J.~Lamblin and P.~Lautridou, \textit{et al.} Nucl. Instrum. Meth. A \textbf{555} (2005), 148.

\bibitem{Anna:2015}
Anna Nelles, Stijn Buitink, Heino Falcke, Jörg Hörandel, Tim Huege, and Pim Schellart.  Astropart. Phys., \textbf{60} 13–24, (2015).

\bibitem{Coreas2013}
T. Huege, M. Ludwig and C.W. James, AIP Conf. Proc. 1535 (2013), 128.


\bibitem{PierreAuger:2014ldh}
A.~Aab \textit{et al.} [Pierre Auger], Phys. Rev. D \textbf{89} no.5, 052002 (2014).


\bibitem{Glaser:2016}
C. Glaser, M. Erdmann, J.R. H{\"o}randel, T. Huege, J. Schulz, JCAP \textbf{09}, 024 (2016)

\bibitem{Welling:2019scz}
C.~Welling, C.~Glaser and A.~Nelles, JCAP \textbf{10}, 075 (2019). 

\bibitem{maris:2013}
I. Maris. [Pierre Auger], PoS EPS-HEP2013, 405 (2013).

\bibitem{PierreAuger:2015hbf}
A.~Aab \textit{et al.} [Pierre Auger], Phys. Rev. D \textbf{93}, no.12, 122005 (2016).

\bibitem{Dembi:2016}
H. P. Dembinski, B. K\'egl, I. C. Mari\c{s}, M. Roth, and D. Veberi\v{c}, Astropart. Phys. \textbf{73}, 44 (2016).

\bibitem{Gottowik:2017wio}
M.~Gottowik, C.~Glaser, T.~Huege and J.~Rautenberg, Astropart. Phys. \textbf{103} (2018), 87-93.

\bibitem{Schroeder:2024bzf}
F.~G.~Schroeder \textit{et al.} [IceCube], PoS ARENA2024 (2024), 034.

\bibitem{Glaser:2022lky}
C.~Glaser, S.~McAleer, S.~Stj\"arnholm, P.~Baldi and S.~W.~Barwick, Astropart. Phys. \textbf{145} (2023), 102781

\bibitem{Buitink_2014}
S. Buitink \textit{et al.}, Phys. Rev. D {\bfseries 90}, 082003 (2014)



\bibitem{decoene2023radio}
V.~Decoene, O.~Martineau-Huynh, and M.~Tueros,
Astropart. Phys., vol. \textbf{145}, p.~102779, (2023).

\bibitem{iminuit}
H. Dembinski, P. Ongmongkolkul \textit{et al.}, scikit-hep/iminuit, 10.5281/zenodo.3949207 (2020).

\bibitem{nelles2015radio}
A. Nelles \textit{et al.}, JCAP \textbf{05}, 018 (2015).

\bibitem{decoene2021reconstruction}
V.~Decoene, O.~Martineau-Huynh, M.~Tueros, and S.~Chiche,
``A reconstruction procedure for near-horizon extensive air showers based on radio signals,''  
arXiv:2107.03206 [astro-ph.IM], 2021.
\end{thebibliography}

\end{document}